# The Novelty of Syntheses & Varied Applications of ZnO nano systems

**Basavaraj S. Devaramani [1,#], Ramaswamy Y.S. [2], Babu A. Manjasetty[3] and T.R. Gopalakrishnan Nair[4]***

[1]M.Tech. Student (Electrical and Electronics), Research & Industry Incubation Center (RIIC), Dayananda Sagar Institutions(DSI), Bangalore. email: basavaraj@jncasr.ac.in

[#]Honeywell Technologies, Bangalore.

[2]Nano Technology Initiative, RIIC, DSI, Bangalore. email: euronicnano@gmail.com

[3]Proteomics and Bioinformatics Platform, RIIC, DSI, Bangalore. email:babu.manjasetty@gmail.com

[4]RIIC, DSI, Bangalore. email: trgnair@yahoo.com

## Abstract

Controlled synthesis of nano materials is critical for the development of nanotechnologies concerned with respect to shape and size. Especially, ZnO nanosystems, such as wires, belts, needles and films can be easily formed by either physical or chemical approaches such as, Chemical Vapour Deposition, Metal Organic Chemical Vapour Deposition, Pulsed Laser Deposition, Molecular Beam Epitaxy, Sol Gel process, Thermal Annealing method and Solvothermal Oxidation. ZnO has polar surfaces that help in the formation of a wide range of nanostructures such as needles,films rings, springs, bows and helices.

The research on nanostructures has rapidly expanded because of their unique and novel applications in optics, optoelectronics, catalysis, biological sciences and piezoelectricity. Particularly, the ZnO nanosystems have exciting applications owing to their polymorphological structures. This wide-band gap energy of 3.37eV semiconductor enables huge potential for electronic and optical applications. It has unique piezoelectric properties that are very essential for fabricating devices or to enhance the performance of electromechanical devices. It is a biodegradable material suitable for medical and biological applications.

In our work, ZnO nanorods are prepared under specific growth conditions (a mixture of Zinc Acetate and Sodium Hydroxide pellets in the ratio 1:25, favored by Hydrothermal Oxidation of zinc metal at 120°C for 24 hours). ZnO nanorods were grown in the size range 50-150nm. The Scanning Tunneling Microscope (STM) images of the ZnO nanorods display a random distribution of sizes. The X-ray Diffraction (XRD) spectrum reveals a crystalline wurtzite structure for ZnO nanorods.

The advances in synthetic techniques of ZnO nano systems along with its versatile applications have been reported.

## 1. Introduction

For over a century, chemists have developed the ability to control the arrangement of small numbers of atoms inside molecules, leading to revolutions in drug design, plastics, and many other areas. Anotechnology revolves around the Research and Development in the length scale of 0.1 nm to 100 nm to create unique structures, devices, and systems. Many existing technologies depend crucially on processes that take place on the nanometer scale. The surface to volume ratio enhancement of nano forms as compared to bulk counterparts is the crux of bio sensors and nanostructured catalysts;

The several novel applications[1] of nanotechnology include giant magnetoresistance in nanocrystalline materials, nanolayers with selective optical barriers, advanced drug delivery systems; new generation of lasers; systems on a chip; ink jet systems; molecular sieves and many more.

Zinc Oxide (ZnO) has received considerable attention because of its unique optical, semiconducting, piezoelectric, and magnetic properties. ZnO nanostructures exhibit interesting properties including high catalytic efficiency and strong adsorption ability.

The multiplicity of morphologies [2] such as belts, ribbons, cables, rods, tubes, rings, springs, helices, bows, tetrapods, spirals, needles and films are the speciality of ZnO nano systems that forms the basis of its versatile applications.

This wide band-gap (energy 3.37 eV) semiconductor enables huge potential for performance of Micro Electro Mechanical Devices [4]

Recently, the interest has been focused toward the application of ZnO in biosensing because of its biocompatibility [5] and fast electron transfer kinetics.

The polarity of ZnO surfaces help in the formation of a wide range of nano structures such as rings springs bows and helices [6,]

The physical properties of semiconducting nano crystallites is dominated by the spatial confinements of electronic and vibrational excitations. With decrease in size , the gap between the Highest Occupied Molecular Orbital (HOMO) and the Lowest Unoccupied Molecular Orbital (LUMO) widens[.7] It is also possible to fine tune the crucial band gap of ZnO by doping with divalent $Mn^{2+}$, $Ni^{2+}$or Co $^{2+}$ metal ions [8]

An overview of the state-of-the-art of the synthetic methodologies along with our initial research results on the production of ZnO nanorods via ***solvothermal oxidation*** is reported**.**

## 2. The state-of-the art of synthetic strategies

Controlled synthesis of nano materials is critical for the development of nanotechnologies concerned. with respect to shape and size. Several synthetic methods have been exploited to grow varieties of ZnO nanosystems. An assortment of these nano structures have been grown via a variety of methods using suitable substrates like Quartz ,Silica, Si, Ga- As, Li-Nb O3, Li Ta O3,SiC,Mg ZnO, Sapphire, GaN, ZnO or Au , as mentioned below.

a) Chemical Vapour Deposition (CVD)
b) Metal Organic Chemical Vapour Deposition (MOCVD)[9,10]
c) Pulsed Laser Deposition (PLD)[11]
d) Molecular Beam Epitaxy (MBE)[12]

e) Sol-gel process[14]
f) Thermal Annealing method[15].
g) Hydrothermal Oxidation

The fabrication of nano ZnO structures generally comprise of two steps involving
1) Loading the substrate into the combustion chamber prior set to controllable Pressure, (0.1-2 Torr) ,Temperature, (573K) Catalyst and Oxygen flow parameters and
2) Evolution of ZnO nanoforms on the catalytic substrate . NiO 16 or 30-50 Angstrom layer Au deposited by direct current sputtering act as catalysts 17
CVD is a relatively simple method that involves vaporization of Zn powder in oxygen atmosphere at around 1400°C. Vapour pressures of Zn and oxygen as well as reactor temperature are critical for obtaining the desired nanostructures.[18] The reaction between the ZnO and graphite results in carbothermal reduction leading to *in situ* generation of Zn vapors that react with CO or $CO_2$ to form ZnO nanocrystals. The presence of graphite significantly lowers the decomposition temperature to 800-1100°C. Well aligned ZnO nanowires have been conveniently obtained by carbon assisted thermal reduction method (ref 19-22)
In MOCVD method, Edward Frankland's organometallic compound, Zinc diethyl is made to decompose under oxygen or nitrous oxide flow[19-22]
For instance, in a typical MOCVD, a steady increase in temperature of the combustion chamber has resulted in the formation of nanostructures with decreasing diameters [10].
In Pulsed Laser Deposition (PLD) method, method, the laser power has been utilized as a growth parameter to control the diameter of nanorods by controlling the dimension of 3D nucleation[11] An array of ZnO nanorods were synthesized on Silicon substrate at 873 K by 193 nm pulsed laser ablation of a ZnO target in low pressures of oxygen. [23]
Sol-gel method [24] involves typically the reduction of a precursor like Zinc Nitrate hexa hydrate using glucose, when ZnO nano particles in the size range 40-100nm get embedded in 3-5 μ m polycrystalline fibres [25]. ZnO nano particles of average size of 10-15nm homogeneously dispersed in Silica matrix were prepared via sol-gel process by the segregation of Zinc Citrate complex followed by Thermal treatment up to 1073 K [26]
Sol-gel process has the distinct advantage over other methods because of process simplicity and ease of control of film composition.

**Hydrothermal Oxidation**

A novel wet-chemical approach was adopted by Bin Liu et.al, [27] by Hydrothermal Oxidation at 180 °C for the synthesis of monodispersed ZnO nanorods with high single-crystallinity. The method has successfully brought the ZnO nanorod diameter from 150 nm down to the 50 nm range.

Huiying et.al [28] have synthesized ZnO nanorods by hydrothermal route using $Zn(OH)_4^{2-}$ precursor in alcohol solution at 110 °C.
A new simple low temperature thermal route was put forward to synthesise uniform ZnO nano rods in the size range 70nm width and 2μ m length, by using hydrazine hydrate as a mineraliser at 150°C for 8h.. Dewei et.al [29] have synthesized single-crystalline $Pb^{2+}$ doped ZnO nanorods by hydrothermal method in the presence of cetyltrimethyl ammonium bromide (CTAB). The obtained $Pb^{2+}$ doped ZnO nanorods were in diameters of 150 nm and in lengths of 3 μm.

Le, Hong Quang [30] deposited ZnO nano rods of length 1-1.5 μm on p-GaN by low temperature hydrothermal synthesis at 373 K.

Continuous production of highly crystalline ZnO nanorods by supercritical hydrothermal synthesis was reported by _atoshi Ohara et. al. [31] Zinc nitrate aqueous solution was pressurized to 30 MPa at room temperature and rapidly heated to 673 K by mixing with supercritical water and then fed into a tubular reactor for a residence time of about 10 s. Surface modification of ZnO nanorods with organic reagents by the supercritical hydrothermal synthesis was also made by the authors.

These synthetic strategies (a-g) constitute the basis for developing versatile applications of ZnO nano forms.

## 3. Nanotechnology at DSI

Our research focuses on the fabrication of perfectly aligned and uniform ZnO nanorods which can be scaled up to practical applications. In this direction, we have recently prepared ZnO nanorods in the size range 50- 150nm, under specific growth conditions .

## 4. Experimental Methodology

Analar grade Zinc acetate dihydrate (molar mass 219.4) was mixed with Analar grade Sodium Hydroxide (molar mass 40) in the molar ratio 1:25 and heated separately in aqueous or ethanol media to 393K for 24 h. Hydrothermal Oxidation favoured the evolution of ZnO nano rods

## 5. Characterisation : Results and Discussion

The Scanning Tunneling Microscope (STM) images of these ZnO nanorods display a random distribution of sizes (Figures 1- 4). The X-ray Diffraction (XRD) peaks are quite sharp indicating the crystalline nature of the particles. The XRD patterns of the products obtained by aqueous and ethanol mediated thermal routes are very similar and match the standard file of ZnO (JCPDS Card No. 79-0206). [32-33]

The patterns recorded over a range of angle 2 theta values from 30° to 75° reveals a crystalline wurtzite structure [3] (Figures 1- 4 ).

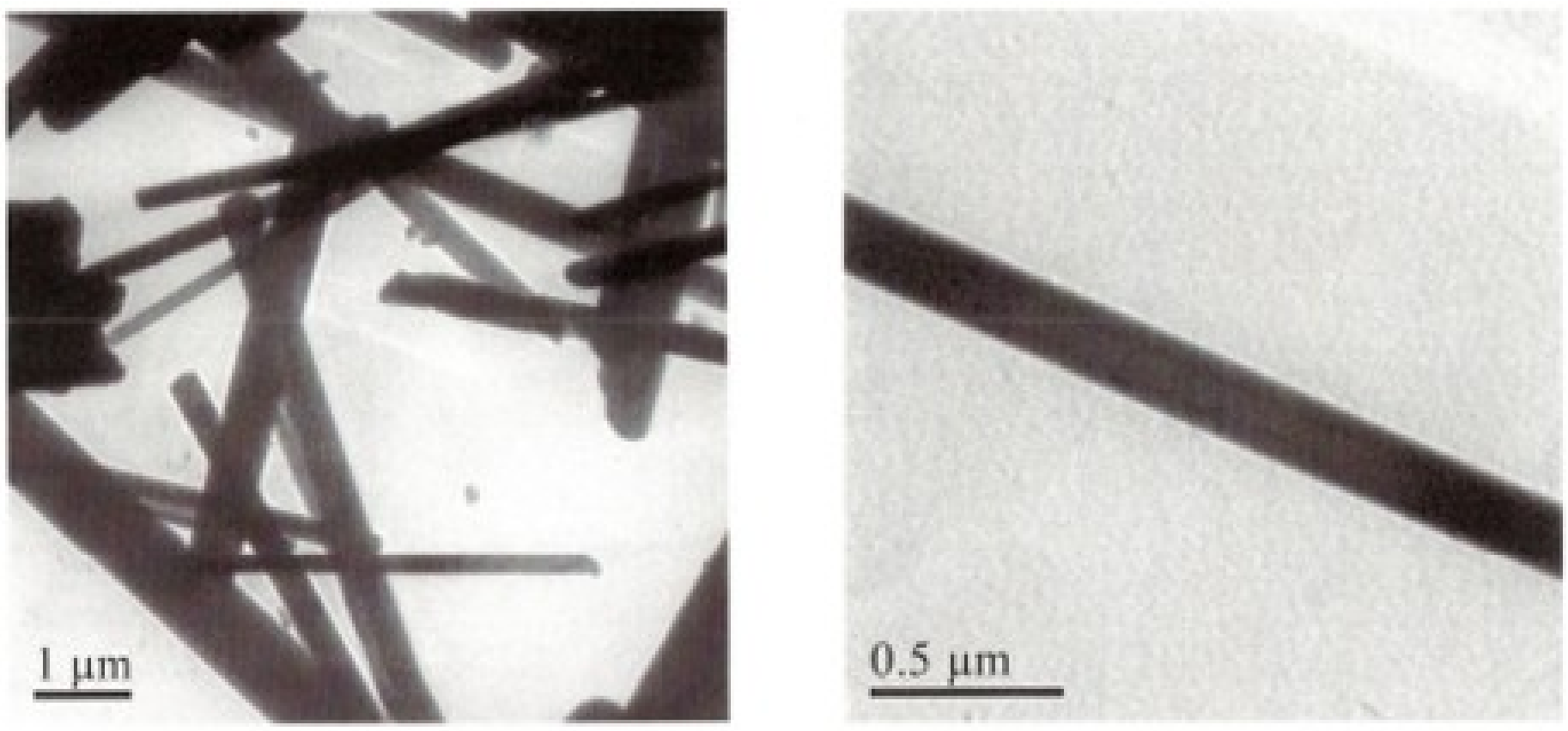

**Figures 1 4. Distribution of ZnO nanorods are shown (size varies from 50 to 150nm)**

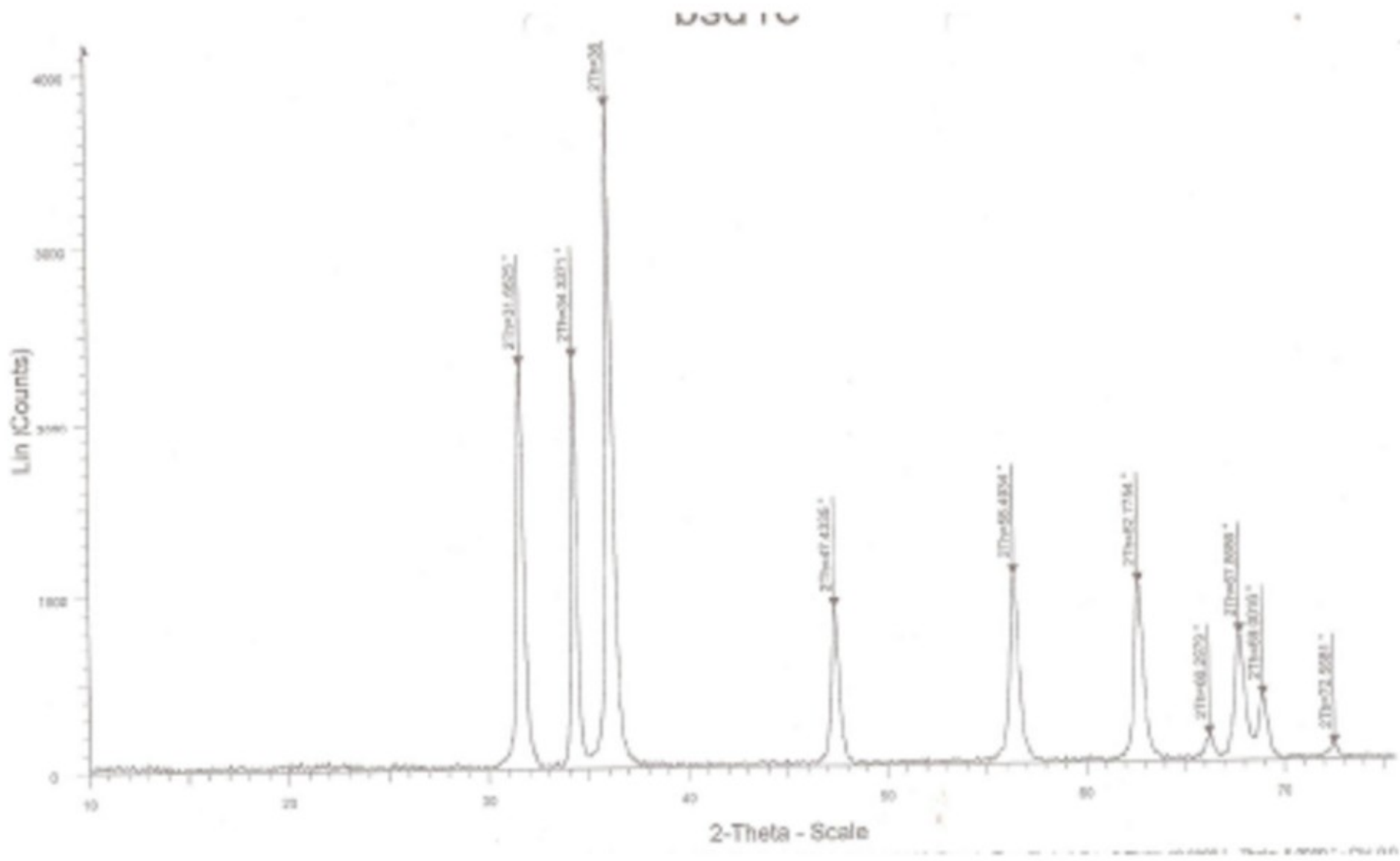

Figures 5.-6 Typical XRD pattern of ZnO nanorods.

Doping with suitable metal ions of ZnO nanoforms to manipulate the electronic properties will be undertaken. Development of indigenous metal catalysts coatings (Au thin film) on substrates (Si, Quartz and Sapphire) suitable for electronic devices is our extended area of research.

## 6. Conclusions.

Nanotechnology is a novel branch of futuristic science and engineering. The synthetic studies of multiple morphologies of ZnO structures constitute the basis for developing versatile applications in the development of new domains.

ZnO has been recognized as one of the key materials in Oxide-Electronics[34] enabling the realization of technologies of ultraviolet light emitters, detectors, thin film transistors, spintronics, self organized nanostructures, and so on. ZnO nanowires are extremely sensitive even to tiny forces in the nano- to pico-newton range. This principle is used to make ZnO pressure sensors that can be implanted in the body owing to their biocompatibility .